\documentclass[prb,reprint,superscriptaddress,showpacs]{revtex4-1}
\usepackage{graphicx}
\usepackage{amssymb}
\usepackage{amsmath}
\usepackage{bbold}
\usepackage{url}
\usepackage{dcolumn}
\usepackage{prettyref}
	\newcommand{\pr}[1]{\prettyref{#1}}
	\newrefformat{app}{appendix~\ref{#1}}
	\newrefformat{eqn}{Eqn.~(\ref{#1})}
	\newrefformat{eqns}{Eqns.~(\ref{#1})}
	\newrefformat{sec}{section~\ref{#1}}
	\newrefformat{subsec}{section~\ref{#1}}
	\newrefformat{subsubsec}{section~\ref{#1}}
	\newrefformat{fig}{Fig.~\ref{#1}}
\usepackage[breaklinks,bookmarksnumbered]{hyperref}
\newcommand{\eps}{\epsilon}
\newcommand{\V}[1]{\mathbf {#1}}
\newcommand{\EQselfenergyDetailed}[2]{
\raisebox{#1}{
	\includegraphics[width=#2]{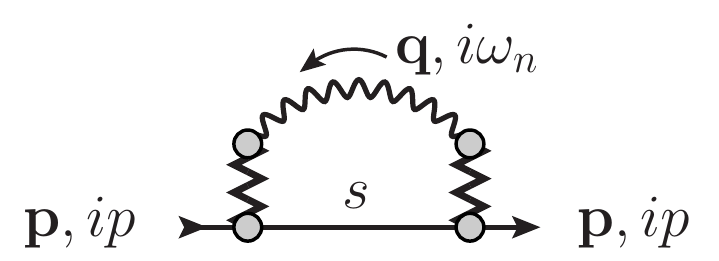}
}
}
\usepackage[usenames,dvipsnames]{color}
\begin{document}
\title{Temperature dependence of the conductivity of graphene on boron nitride}
\author{J{\"u}rgen Schiefele}
\affiliation{%
Departamento de F\'isica de Materiales,
Universidad Complutense de Madrid,
E-28\,040, Madrid, Spain%
}
\author{Fernando Sols}
\affiliation{%
Departamento de F\'isica de Materiales,
Universidad Complutense de Madrid,
E-28\,040, Madrid, Spain%
}
\author{Francisco Guinea}
\affiliation{%
Instituto de Ciencia de Materiales de Madrid,
CSIC,
Cantoblanco,
E-28\,049, Madrid, Spain
}
%
%
%
\date{May 1, 2012}
%
%
%
%
%
%
\begin{abstract}
The substrate material of monolayer graphene influences the charge carrier mobility
by various mechanisms. 
At room temperature, the scattering of conduction electrons
by phonon modes localized at the substrate surface can severely limit the
charge carrier mobility.
We here show that for substrates made of the piezoelectric hexagonal boron nitride (hBN), in
comparison to the widely used SiO$_2$, this mechanism of remote phonon scattering is
--at room temperature-- weaker by almost an order of magnitude, and causes a resistivity of
approximately 3\,$\Omega$.
This makes hBN an excellent candidate material for future graphene based electronic devices operating at room temperature.
\end{abstract}
\maketitle
%
%
%
%
%
%
%
%
%
%
%
%
%
%
\section{Introduction}
Apart from the great theoretical interest in the electronic properties of graphene\cite{Guinea_2009}
spurred by the experimental realization of these one atom thick carbon layers,\cite{Novoselov_2004}
the new material  has been soon recognized as a promising candidate for technological applications.\cite{Geim_2007}
At low temperatures, electrons in free standing (that is, suspended above the substrate)
graphene sheets can travel in an essentially ballistic manner,
without scattering over distances of the order of a micron.\cite{Du_2008}
With the carrier density tuned to finite values by an  applied  gate voltage,
the temperature dependence of the resistivity in suspended graphene sheets
shows a metallic behavior: at temperatures above 50\,K, it increases  linearly 
with temperature. 
The carrier mobility at room temperature in suspended graphene is severely limited by scattering by out of plane flexural phonons%
,\cite{Cetal10,Ochoa_2010}
resulting in a resistivity that increases quadratically with the temperature.
The effect of flexural phonons on the charge carrier mobility  can be eliminated by
applying strain or placing graphene on a substrate.\cite{Nika_2012}

Compared to graphene devices on substrates, suspended graphene samples offer charge carrier mobilities which are higher by a
factor of ten, approximately.\cite{Tan_2007}
Likewise, the thermal conductivity of graphene 
on a substrate can be an order of magnitude lower than that of suspended graphene,
due to damping of the flexural phonon modes.\cite{Balandin_2011,Lindsay_2010,Ong_2011}
For technological applications, however, the suspended geometry imposes rather strict limitations on the device architecture.
Moreover, free floating graphene is always crumpled,\cite{Meyer_2007}
with the corrugations inducing effective fields which in turn influence the charge carriers.\cite{Vozmediano_2010} 
While many experiments so far involved graphene samples deposited over SiO$_2$ or grown over SiC substrates,
hexagonal boron nitride (hBN) has developed into a promising candidate as a substrate material for improved
graphene based devices.\cite{Dean_2010,Mayorov_2011}
hBN is a piezoelectric, large band gap insulator isomorphic to graphite.
Boron and nitrogen atoms occupy the inequivalent sublattices in the Bernal structure.
The lattice mismatch with graphite is small (1.7\%),\cite{Giovannetti_2007} and hBN is expected
to be free of dangling bonds and surface charge traps.

A number of mechanisms limit the low temperature carrier mobility in graphene.\cite{P10,SAHR11,Das_Sarma_2011}
At room temperature, interaction  with optical surface phonon modes on 
the interface between graphene and the SiO$_2$ substrate 
was found to play an important role.\cite{Chen_2008}
It is this mechanism of remote-phonon scattering 
from an hBN substrate
that we want to consider in the present work.
In general,
the polar phonon modes on the surface of the substrate (with energies of $50-200$\,meV) create a long-range electric field,
which influences the electrons in the graphene sheet, typically around 4 \AA{} away.
The influence of remote phonon scattering on the carrier mobility
in two-dimensional electron systems is a well-known phenomenon in semiconductor physics,
and was investigated for quantum wells and other heterostructures including
metal-oxide-semiconductor field-effect transistors
(MOSFETs).\cite{Fischetti_2001,Mori_1989}
The effect is more pronounced in graphene due to the much smaller vertical dimension of the devices,
as determined by the van der Waals distance.
Also, the band gap in semiconductor systems prevents low energy interband  transitions,
which are present in graphene with the two bands touching at the Dirac point or with the metallic band in the case of doping.
For single-layer graphene sheets on substrates of SiC or SiO$_2$, surface-phonon scattering
has been investigated in Refs.~\onlinecite{Konar_2010, Fratini_2008,Li_2010}, employing various methods for calculating
the conductivity  and describing the screening of the interaction by conduction electrons.
In the present paper we calculate the resistivity due to surface phonon scattering for hBN substrates.
We show below that the temperature dependence of this mechanism scales with the thermal population
of surface phonon modes. While for SiO$_2$ substrates, the resistivity due to remote phonon scattering
is known to be comparable or might surpass that due to graphene intrinsic phonons,\cite{Chen_2008,Hwang_2007b}
we show that for hBN, the effect is almost an order of magnitude smaller,
resulting in desirable high charge carrier mobilities.
\section{Electron self energy due to remote phonon coupling}
\label{sec:self_energy}
%
%
%
\subsection{Surface phonons}
\label{subsec:surface_phonons}
The dielectric function of a substrate material with transverse optical modes $\omega_{TO}^{i}$
reads
\begin{equation}
\eps(\omega)
	=
	\eps_{\infty} 
	+
	\sum_i
	f_i \frac{(\omega_{TO}^{i})^2} {(\omega_{TO}^{i})^2 - \omega^2}
\nonumber
\;,
\end{equation}
where 
$\eps_\infty$ denotes the high frequency dielectric constant of the material
and the dimensionless oscillator strengths $f_i$ measure the contribution
of each mode to the screening properties of the material.
They are determined from experimental data by defining intermediate
dielectric constants $\epsilon_i$. These are evaluated at a frequency
just above the corresponding resonance $\omega_{TO}^{i}$
(see for example Ref.~\onlinecite{Fischetti_2001}).
With $\eps_0$ the static dielectric constant, the oscillator strengths are given by
$%
f_i
	=
	\eps_{i-1} - \epsilon_i%
$.
The frequencies $\omega_0^{i}$ of the corresponding surface modes
are determined from the equation\cite{Fratini_2008, Fischetti_2001}
\begin{align}
\epsilon(\omega) + 1
	= 0
\;,
\nonumber
\end{align}
where 1 is the dielectric constant of air and we neglected the dielectric response
of the (atomically thin) graphene layer.

The polar surface modes on the substrate interface create a polarization field
which decays exponentially with the distance from the interface,
and is felt by the electrons in the graphene sheet.
This remote interaction can be brought to the form\cite{Wang_1972, Mori_1989}
\begin{align}
H_{ep}
	=
	\sum_i
	\sum_{\V k, \V q}
	M_q^i
	(a^\dagger_{\V {k+q}} a_{\V k} + b^\dagger_{\V {k+q}} b_{\V k})
	(c^i_{\V q} + c^{i \dagger}_{- \V q})
\;,
\nonumber
\end{align}
where $i$ runs over the different surface modes $\omega_0^i$,
$\V k$ and $\V q$ are two-dimensional momentum vectors parallel to the graphene-substrate interface,
$c^i_{\V q}$  and $c^{i \dagger}_{- \V q}$ are destruction and creation
operators for surface phonons
and the $a_{\V k}$ and $b_{\V k}$ operators are the destruction operators
for electrons on the $A$ and $B$ sublattices of the graphene sheet.
They are coupled by the interaction matrix element
\begin{align}
M_q^i
	&=
	\hbar \omega_0^{i}
	\sqrt{\gamma^{i} \, g^i_{q}}\,
	e^{-q z}
\;,
\label{eqn:M_def}
\end{align}
where $z$ denotes the (positive) distance between the substrate and the graphene sheet,
the coupling strength to the individual surface modes is given by
\begin{align}
\gamma^{i}
	=
	\biggl(
	\frac 1 {\eps_i +1}
	-
	\frac 1 {\eps_{i-1} + 1}
	\biggr)
\;,
\label{eqn:gamma_c}
\end{align}
and the dimensionless function $g^i_q$ reads
\begin{align}
g_{q}^i
	=
	\frac
	{e^2}
	{2 A \eps_{{\rm vac}} \hbar \omega_0^i\,
	(q+ q_{TF})}
\;,
\label{eqn:g_def}
\end{align}
with $A$ denoting the surface area of the interface,
$e$ the electron charge,
and $\eps_{{\rm vac}}$ the permittivity of free space.
The inverse Thomas-Fermi screening length $q_{TF}= e^2 E_F / (\pi \epsilon_{{\rm vac}} \hbar^2 v_F^2)$
incorporates the effect of dynamic screening from the conduction electrons.
\begin{table}
\caption{Material parameters for hBN, SiO$_2$ and SiC
substrates, taken from Refs.~\onlinecite{Geick_1966}, \onlinecite{Fischetti_2001}, and \onlinecite{Nienhaus_1989}, respectively.
The bulk optical phonon frequencies $\omega_{TO}$ are given at the $\Gamma$ point.
For materials with two surface modes, $\epsilon_\infty$ figures as the intermediate dielectric
constant $\epsilon_2$ in the calculation of the coupling parameter $\gamma^{2}$
in \pr{eqn:gamma_c}.
}
\begin{ruledtabular}
\begin{tabular}{l ccc}
\,	&  {hBN}   & {SiO$_2$}  & {SiC}\\
$\epsilon_0$  & 5.09 & 3.90 & 9.7 \\
$\epsilon_1$  & 4.57 & 3.05 & \, \\
$\epsilon_\infty$  & 4.10 & 2.50 & 6.5 \\
$\hbar \omega_{TO}^{1}$ [meV] & 97.4 & 55.6 & 97.1 \\
$\hbar \omega_{TO}^{2}$ [meV] & 187.9 & 138.1 & \, \\
$\gamma^{1}$  & 0.0153 & 0.0428 & 0.040 \\
$\gamma^{2}$  & 0.0165 & 0.0388 & \, \\
$\hbar \omega_0^{1}$ [meV] & 101.6 & 61.0 & 116 \\
$\hbar \omega_0^{2}$ [meV] & 195.7 & 149.0 & \, \\
$z$ [\AA] & 3.4 & 4.0 & 4.0 \\
\end{tabular}
\end{ruledtabular}
\label{table_parameters}
\end{table}
%
%
%
%
%
%
%
%
\subsection{Electron self energy}
In considering the charge carriers in graphene, we limit
our treatment to the Dirac cone approximation, where the dispersion
reads
\begin{align}
E_s (\V p)
	=
	s v_F |\V p|
\;,
\label{eqn:dispersion}
\end{align}
with $s=1$ denoting the conduction ($\pi^*$) band and $s=-1$ the valence
($\pi$) band,
and the Fermi velocity $v_F\approx 10^6$\,m/s.
In leading order perturbation theory, the self energy acquired by these
Dirac fermions due to coupling to a remote substrate phonon is
\begin{align}
\Sigma_s & (\V p, ip)
	=
	\EQselfenergyDetailed{-1.5ex}{0.2\textwidth}
\label{eqn:se1}
\\[1.5ex]
	&=
	- k_B T
	\sum_{i \omega_n, \V q}
	M_q^2 D^{(0)}(i \omega_n)
	G^{(0)}_s (\V p + \V q, ip + i \omega_n)
\;,
\nonumber
\end{align}
where $D^{(0)}$ and $G^{(0)}$ denote the free thermal (Matsubara) Green's function
for the phonon and the electron, respectively, and the interaction (\ref{eqn:M_def})
is depicted by the vertical zigzag lines in the diagram.
(See  Eqns.~(\ref{eqn:D0_def}) and (\ref{eqn:G0_def}) in the Appendix and
Refs.~\onlinecite{Mahan_old, Fetter} for details.)
After summation over the bosonic Matsubara frequency $\omega_n$,
a shift of the integration variable to $\V k = \V p + \V q$,
and rotating back to real frequencies ($i p \to \omega - \mu + i \eta$),
we are left with
%
\begin{align}
\Sigma_s^{i} &(\V p, \omega)
	=
	\sum_{\V k}
	\,
	\biggl\{
	\frac
	{N_0^i + 1 - n_F[E_s(k)]}
	{\hbar(\omega -\mu - \omega_0^{i}) - E_s(k) + i \eta}
\nonumber
\\
	&+
	\frac
	{N_0^i +  n_F[E_s(k)]}
	{\hbar(\omega -\mu + \omega_0^{i}) - E_s(k) + i \eta}
	\biggr\}
	\,
	\mathcal{G}_s (\V k)
	\bigl(M^i_{\V p - \V k}\bigr)^2
\label{eqn:se2}
\end{align}
Here,
$n_F$ denotes the Fermi distribution,
$N_0^i$
the thermal occupation of the phonon,
and the appearance of the $2\times2$ matrix $\mathcal{G}_s $ in the electron Green's
function (see \pr{eqn:Gs_def})
is due to the spinor representation of the electron wavefunction in graphene.
%
%
%
\section{Quasiparticle scattering rate}
\label{sec:rate}
The scattering rate can be obtained from \pr{eqn:se2} by multiplying
the self energy from the left and right with spinor wavefunctions $\V{F}_{\pm 1}$
(see \pr{eqn:F_def})
for the ingoing and outcomming electron, respectively, and afterwards taking the imaginary part:
\begin{align}
\Gamma^{i} (\omega 
	)
	&=
	- \frac 1 {2 \hbar} \operatorname {Im} \,
	\sum_{s=\pm 1}
	\V{F}^\dagger_{\operatorname{sgn}[\omega]}(\V p)
	\Sigma_{s}^{i} (\V p, \omega)
	\V{F}_{\operatorname{sgn}[\omega]}(\V p)
\;,
\label{eqn:rate1}
\end{align}
where the incoming momentum $\V p$ is set on-shell,
that is $\hbar \omega = \pm \hbar v_F |\V p| - \mu$,
and $\operatorname{sgn}[\omega]=\pm 1$ for $\omega>0$ ($\omega<0$),
respectively.
The imaginary part is obtained via $\operatorname{Im} [x + i\eta]^{-1} = - \pi \delta (x)$,
which yields the following rates $\Gamma^{i}_{\pm}(\omega)$ for emission 
or absorption of a phonon with frequency $\omega_0^{i}$:
\begin{subequations}
\label{eqn:rate2}
\begin{align}
&\Gamma^{i}_+(\omega)
	=
	\frac{\pi} {2 \hbar^2}
	\bigl[N_0 + 1- n_F(\omega-\omega_0^{i}) \bigr]
\label{eqn:rate2_emission}
\\
	&\times
	\sum_{\V k}
	M^2_{\V p - \V k}
	\delta(v_F k - | \omega-\omega^{i}_0|)
	\begin{cases}
	f_{-1}	&	\text{for} \; 0\leq\omega<\omega_0^{i}\\
	f_{1}	&	\text{else}\\
	\end{cases}
\;,
\nonumber
\\[2ex]
&\Gamma^{i}_-(\omega)
	=
	\frac{\pi} {2 \hbar^2}
	\bigl[N_0 + n_F(\omega+\omega_0^{i}) \bigr]
\label{eqn:rate2_absorbtion}
\\
	&\times
	\sum_{\V k}
	M^2_{\V p - \V k}
	\delta(v_F k - | \omega+\omega^{i}_0|)
	\begin{cases}
	f_{-1}	&	\text{for} \, -\omega_0^{i} \leq \omega<0\\
	f_{1}	&	\text{else}\\
	\end{cases}
.
\nonumber
\end{align}
\end{subequations}
The total scattering rate is obtained by summing the absorption
and emission rates for scattering from all surface phonons $i$.
The angular factor
\begin{align}
f_{\pm1} (\V k, \V p)
	&=
	\frac 1 2 [1 \pm \operatorname{cos}(\theta_{\V k} -\theta_{\V p})]
\;,
\nonumber
\end{align}
in Eqns. (\ref{eqn:rate2}), where
$\V k = k (\operatorname{cos} \theta_{\V k}, \operatorname{sin} \theta_{\V k})$,
distinguishes between intraband (upper sign) and interband (lower sign) scattering%
\footnote{%
Eqns. (\ref{eqn:rate2}) are identical to the result obtained  in 
Ref.~\protect \onlinecite{Fratini_2008},
while in Refs.~\protect \onlinecite{Konar_2010} and \protect \onlinecite{Li_2010}, interband scattering is neglected.}%
,
see \pr{eqn:f_angular} and the inset in \pr{fig:GammaVsOmega}.
Terms describing interband scattering  are seen to contribute only in the
range $|\omega| \le \omega_0^{i}$.

\begin{figure}
\centering
\includegraphics[width=\columnwidth]{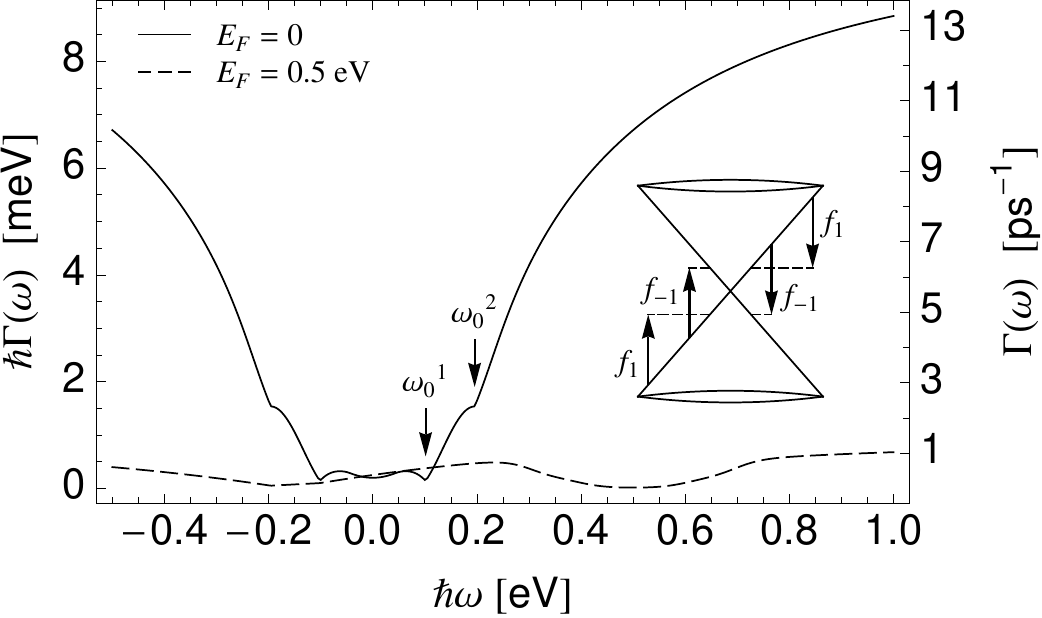}
\caption{%
Quasiparticle scattering rate $\Gamma(\omega)$ in single layer graphene
due to  surface phonons of the hBN substrate at $T=300K$
[see eqns.~(\ref{eqn:rate2})].
The frequencies of the optical surface modes $\omega_0^{1,2}$ are
given in table \ref{table_parameters},
the distance between graphene layer and substrate is 3.40\,\AA.
Full line:
Intrinsic graphene ($E_F=0$).
Dashed line:
Extrinsic graphene with $E_F=0.5$\,eV.
Inset:
The angular factor $f_{\pm 1}$ [see \pr{eqn:f_angular}] in the scattering rate
distinguishes between interband and intraband scattering.
}
\label{fig:GammaVsOmega}
\end{figure}

Figure \ref{fig:GammaVsOmega} shows $\Gamma(\omega)$ at $T=300$\,K for
intrinsic graphene ($E_F=0$, full line) and extrinsic graphene with $E_F=0.5$\,eV
(corresponding to a carrier density of $n=1.83\cdot10^{13}$\,cm${^{-2}}$,
dashed line)
on a hBN substrate.
The distance between graphene layer and substrate is set to 3.40\,\AA,
as found in Ref.~\onlinecite{Giovannetti_2007}
for a stacking configuration with one carbon over N, and the
other carbon centered above a hBN hexagon\footnote{For other possible stacking configurations, this distance can be around 0.2\,\AA{}
smaller or larger, see Ref.~\protect \onlinecite{Giovannetti_2007}.}.
For electron energies around $E_F \pm \hbar \omega_0^{1}$,
scattering is strongly suppressed
and goes to zero at $T=0$,
as there are no empty electronic states below the Fermi energy to scatter into.

The large difference in $\Gamma$ between the doped and undoped case
is due to the Thomas-Fermi screening in \pr{eqn:g_def}.
As this model assumes the instantaneous reaction of the screening
charges, our values for $\Gamma$ at finite doping present a
lower bound on the scattering rate.\cite{Fratini_2008, Wunsch_2006}
If screening is completely neglected, the rate at $E_F = 0.1$\,eV is larger by a factor
of four, approximately, which presents an upper bound for the rate.
%
%
%
%
%
%
%
%
%
%
%
%
%
\begin{figure}
\centering
\includegraphics[width=\columnwidth]{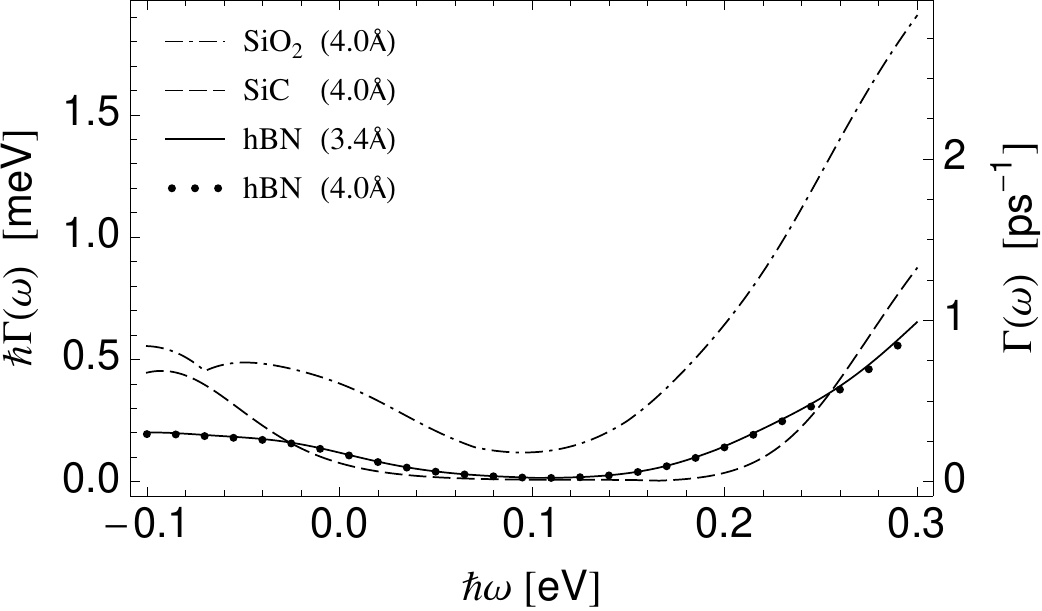}
\caption{%
Scattering rate $\Gamma(\omega)$
at $T=300K$,  $E_F=0.1$\,eV, for different substrate materials.
The plot shows the energy range around the Fermi energy,
which is relevant for the conductivity,
for hBn, SiC and SiO$_2$ substrates.
(See table \ref{table_parameters} for the material parameters.)
The full line is evaluated with the distance between hBN and graphene set to 3.4\,$\AA$ 
(see Ref.~\protect\onlinecite{Giovannetti_2007}).
For comparison, the dots show the rate at a distance of 4\,$\AA$, as for the other two materials 
(see Ref.~\protect\onlinecite{Fratini_2008}).
The resulting relative difference in $\Gamma(E_F)$ is less than 5 per cent.
}
\label{fig:gammaDetail}
\end{figure}
%
%
%
%
%
%
\section{Substrate limited conductivity}
\label{sec:conductivity}
%
%
%
%
%
%
\begin{figure}
\centering
\includegraphics[width=\columnwidth]{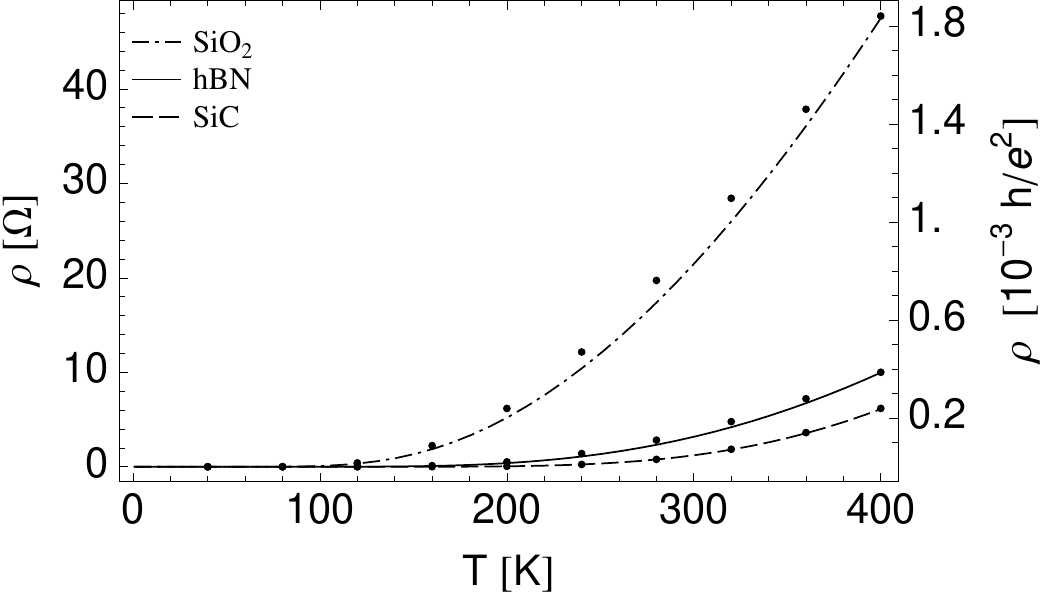}
\caption{%
Temperature dependence of the
resistivity for SiO$_2$, hBN and SiC (top to bottom).
The dots correspond to the population factor
$%
\rho_0 / \operatorname{sinh}[\hbar \omega_0^{1} / (k_B T)]
$
(see \pr{eqn:population}),
with $\rho_0=178, 95, 381 \, \Omega$
for SiO$_2$, hBN, and SiC, respectively.
}
\label{fig:RhoVsT}
\end{figure}
%
%
%
%
\begin{figure}
\centering
\includegraphics[width=\columnwidth]{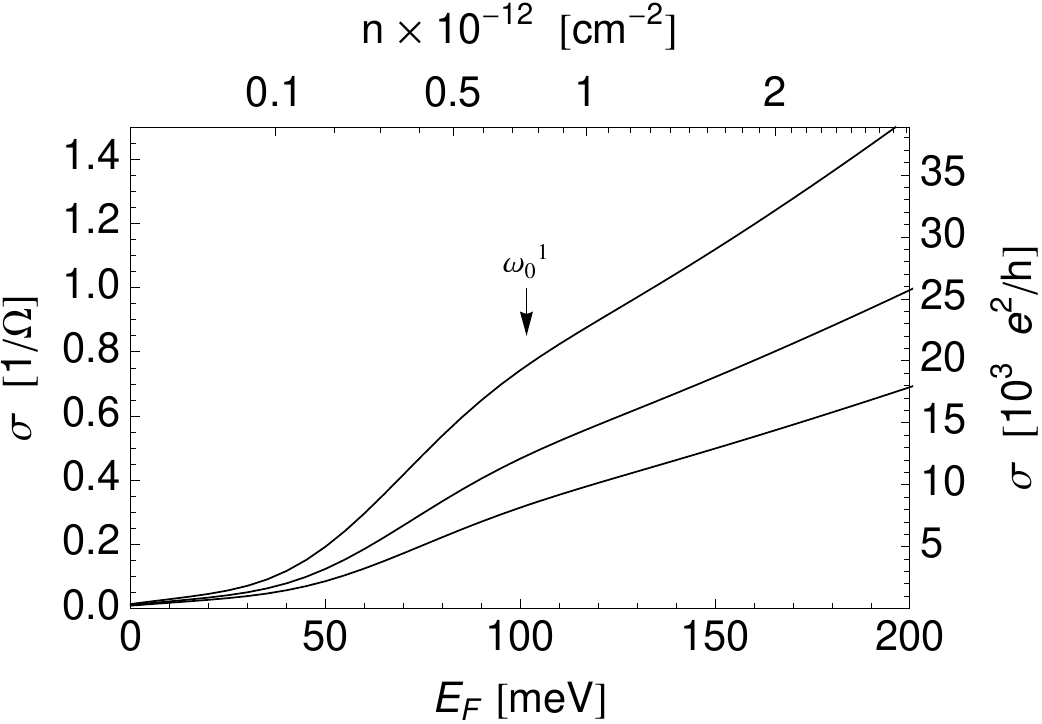}
\caption{%
Conductivity versus Fermi energy for different temperatures $T=300,275,250$\,K (bottom to top).
}
\label{fig:SigmaVsEF_hBN}
\end{figure}
%
%
%
%
%
We calculate the electrical dc conductivity $\sigma$ of the
graphene layer via the Boltzmann equation:
\begin{align}
\sigma
	=
	\frac {e^2} {h}
	\int d\omega \,
	\frac{|\omega|} {\Gamma_{\rm tr}(\omega)}
	\biggl[
	-
	\frac {d \, n_F (\omega)}{d \omega}
	\biggr]
\label{eqn:conductivity1}
\end{align}
Here, the transport scattering rate $\Gamma_{\rm tr}$ is defined as
in \pr{eqn:rate2}, but with an additional factor
$[1 \mp \operatorname{cos}(\theta_{\V k} -\theta_{\V p})]$
in the integrand (with the upper sign for intraband and the lower one for interband scattering),
which lends more weight to large angle scattering events.\cite{Mahan_old}
As the derivative of the Fermi function is sharply peaked around the Fermi energy,
the integrand of \pr{eqn:conductivity1} depends mainly  on the scattering rate
for $\omega \approx E_F/\hbar$,
where both the phonon emission and phonon absorption term in \pr{eqn:rate2} are proportional to
\begin{align}
	\omega_0^{i}
	/
	\operatorname{sinh}[\hbar \omega_0^{i} / (k_B T)]
\;.
\label{eqn:population}
\end{align}
At a given temperature, this is a decaying function of $\omega_0^{i}$.
Thus scattering in the relevant range is larger for substrate materials
with lower phonon frequencies,
which translates into a lower conductivity.
In \pr{fig:gammaDetail}, we compare scattering rates at electron energies
near $E_F=0.1$\,eV for different substrate materials.
For electrons at the Fermi energy, the  rates for SiC and hBN differ roughly by a factor of two,
while the scattering rate on SiO$_2$ is around ten times larger than that on SiC.
As the  resistivity is calculated from \pr{eqn:conductivity1},
its temperature dependence,  shown in \pr{fig:RhoVsT},
is likewise dominated by the population factor (\ref{eqn:population}).
We here assume a temperature which is constant throughout the sample, 
with local heating of the graphene layer prevented by heat dissipation through the substrate.\cite{Sevincli_2011}

Regarding the dependence on the charge carrier concentration,  the conductivity (\ref{eqn:conductivity1}) depends approximately
linear on $E_F$ for energies larger than $\omega_0^1$ (see \pr{fig:SigmaVsEF_hBN}).
This behaviour is intermediate between that typical for
short range scattering, $\sigma(E_F) \approx$ const., and for charged impurity scattering,
$\sigma(E_F) \approx E_F^2$.\cite{Nomura_2007}
%
%
%
%
%
\section{Summary and conclusions}
We have analyzed the scattering of graphene electrons from phonons localized at the substrate
surface.
The temperature dependence of the scattering rate for electrons with energies near $E_F$
is proportional to the thermal population of these surface phonons,
and the same holds for the resistivity. 
The resistivity induced by the polar surface phonons of hBN substrates
at room temperature is of the order of $3 \, \Omega$, 
an order of magnitude smaller than the corresponding resistivity of graphene on
SiO$_2$ substrates, in consistency with the higher frequency of these modes in hBN.

For graphene -- SiO$_2$ devices, the temperature dependence of the conductivity was 
experimentally found to consist of two contributions:\cite{Chen_2008}
scattering by acoustic phonons in graphene  and scattering by surface phonons of the substrate.
The former shows a linear $T$ dependence, is independent of the carrier density, and contributes approximately
30\,$\Omega$ at room temperature.
The latter is carrier density dependent, follows the  bosonic population of the surface modes,
and becomes only relevant for $T>200\,$K. 
At room temperature it is found to dominate the linear term. 
In experiments with monolayer graphene on hBN,\cite{Dean_2010} 
the conductivity shows the same linear temperature dependence as that  reported in Ref.~\onlinecite{Chen_2008}
for SiO$_2$.
No indication of activated remote surface phonon scattering was seen up to the experimentally realized temperatures
of 200\,K.
An estimate of the temperature dependence of the resistivity of graphene on hBN can also be obtained 
from the experiments reported in Ref.~\onlinecite{Mayorov_2011}.
The result is consistent with the assumption that the room temperature 
resistivity of graphene on hBN is mainly determined by in plane phonons.

Apart from electron-phonon scattering, Coulomb scattering by charged impurities gives rise to 
a temperature independent residual resistivity.
At carrier concentrations of $10^{12}$\,cm$^{-2}$, this is reported with approximately $400\,\Omega$
for SiO$_2$  and roughly three times smaller values for hBN.\cite{Chen_2008,Dean_2010} 

Our calculation suggests that for graphene on hBN,
the temperature-dependent part  of the resistivity is
--in contrast to SiO$_2$ substrates--
at room temperature still dominated by electron scattering from the graphene intrinsic phonons,
the contribution from interface phonons being negligible.
Together with the smaller residual resistivity of graphene on hBN as compared to SiO$_2$, 
this might allow for graphene -- hBN  devices with charge carrier mobilities close to that
of  suspended graphene.
%
%
%
\begin{acknowledgements}
We are thankful to  Andre Geim for providing helpful information.
J.S. would like to thank Christopher Gaul and Ivar Zapata for helpful discussions.
The authors acknowledge support from the 
Marie Curie ITN \emph{NanoCTM} 
and from MICINN (Spain) through Grant No. 
FIS2010-21372 
and 
FIS2008-00124.
\end{acknowledgements}
%
%
%
%
\appendix
%
%
%
%
%
\section{Matsubara Green's functions and electronic wavefunctions}
\label{app:Matsubara}
The Fourier transform of the free thermal Green's function for a
surface phonon with frequency $\omega_0$ is given by
\begin{align}
D^{(0)}(i \omega_n)
	&=
	\frac{2 \hbar \omega_0}{(i \omega_n)^2 - (\hbar \omega_0)^2}
\;,
\label{eqn:D0_def}
\end{align}
where the bosonic Matsubara frequencies are defined as
$\omega_n = 2 \pi k_B T \, n$ with integer $n$.\cite{Mahan_old}

The wavefunction for electronic states in graphene near one of the Dirac points is\cite{Ando_2006_b, Guinea_2009}
\begin{align}
\V{F}_{s} (\V r)
	=
	A^{-1/2}\,
	\V{F}_{s} (\V q)
	\operatorname{exp}(i \V q \cdot \V r)
\nonumber
\end{align}
where $s=\pm1$ denotes the band index,
$A$ is the area of the system and
\begin{align}
\V{F}_s (\V q)
	&=
	\frac 1 {\sqrt{2}} \,
	\left( \begin{array}{c}
	e^{-i \theta_{\V q}}
	\\
	s
	\end{array} \right)
\;.
\label{eqn:F_def}
\end{align}
In the same spinor representation,
the electron Green's function is written as the $2 \times 2$ matrix\cite{Stauber_2008}
\begin{align}
G^{(0)}_s (\V k, i \omega_n)
	&=
	\mathcal{G}_s (\V k)
	\frac{1}{i \omega_n -  E_s (\V k)+\mu}
\label{eqn:G0_def}
\end{align}
where
$\omega_n = (2 n + 1) \pi k_B T$
are the fermionic Matsubara frequencies,
$E_s$ is the energy of the electron (see \pr{eqn:dispersion}) within the
Dirac cone approximation and
\begin{align}
\mathcal{G}_s (\V k)
	&=
	\frac 1 2 \,
	\left( \begin{array}{c c}
	1	&	s e^{-i \theta_{\V k}}
	\\
	s e^{i \theta_{\V k}}	& 1
	\end{array} \right)
\;.
\label{eqn:Gs_def}
\end{align}
The angular factor $f_{\pm 1}$ appearing in \pr{eqn:rate2}
is composed of
\begin{align}
\V{F}_{s'}^\dagger  (\V p)
\mathcal{G}_s (\V k)
\V{F}_{s'} (\V p)
	=
	\frac 1 2 [1 + s s' \operatorname{cos}(\theta_{\V k} -\theta_{\V p})]
	\equiv
	f_{s \cdot s'} (\V k, \V p)
\;.
\label{eqn:f_angular}
\end{align}
%
%
%

\end{document}